%% file: eccv2022submission.tex
\documentclass[runningheads]{llncs}
\usepackage{graphicx}

\usepackage{comment}
\usepackage{amsmath,amssymb} 
\usepackage{graphicx}
\usepackage{booktabs}

\input{macros}

\usepackage[pagebackref,breaklinks,colorlinks=True]{hyperref}
\usepackage[capitalize]{cleveref}
\crefname{section}{Sec.}{Secs.}
\Crefname{section}{Section}{Sections}
\Crefname{table}{Table}{Tables}
\crefname{table}{Tab.}{Tabs.}
\usepackage{orcidlink}
\usepackage{multirow} 

\usepackage[accsupp]{axessibility}  

\begin{document}
\pagestyle{headings}
\mainmatter
\def\ECCVSubNumber{4122}
\title{Synthesizing Light Field Video\\ from Monocular Video}
\titlerunning{Synthesizing LF Video from Monocular Video}
\authorrunning{Shrisudhan G. et al.}
\author{
Shrisudhan Govindarajan\orcidlink{0000-0002-3546-8223} \and
Prasan Shedligeri\orcidlink{0000-0002-0342-9393} \and \\
Sarah\orcidlink{0000-0003-1348-7586}  \and
Kaushik Mitra\orcidlink{0000-0001-6747-9050}}

\institute{Indian Institute of Technology Madras, Chennai, India 
}

\maketitle

\begin{abstract}
The hardware challenges associated with \gls{lf} imaging has made it difficult for consumers to access its benefits like applications in post-capture focus and aperture control.
Learning-based techniques which solve the ill-posed problem of \gls{lf} reconstruction from sparse (1, 2 or 4) views have significantly reduced the need for complex hardware.
\gls{lf} \emph{video} reconstruction from sparse views poses a special challenge as acquiring ground-truth for training these models is hard.
Hence, we propose a self-supervised learning-based algorithm for \gls{lf} video reconstruction from monocular videos.
We use self-supervised geometric, photometric and temporal consistency constraints inspired from a recent learning-based technique for \gls{lf} video reconstruction from stereo video.
Additionally, we propose three key techniques that are relevant to our monocular video input.
We propose an explicit disocclusion handling technique that encourages the network to use information from adjacent input temporal frames, for inpainting disoccluded regions in a \gls{lf} frame.
This is crucial for a self-supervised technique as a single input frame does not contain any information about the disoccluded regions.
We also propose an adaptive low-rank representation that provides a significant boost in performance by tailoring the representation to each input scene.
Finally, we propose a novel refinement block that is able to exploit the available \gls{lf} image data using supervised learning to further refine the reconstruction quality.
Our qualitative and quantitative analysis demonstrates the significance of each of the proposed building blocks and also the superior results compared to previous state-of-the-art monocular \gls{lf} reconstruction techniques.
We further validate our algorithm by reconstructing \gls{lf} videos from monocular videos acquired using a commercial GoPro camera. An open-source implementation is also made available\footnote{\scriptsize \url{https://github.com/ShrisudhanG/Synthesizing-Light-Field-Video-from-Monocular-Video}}.

\end{abstract}

\begin{figure}[t]
    \centering
    \includegraphics[width=\columnwidth]{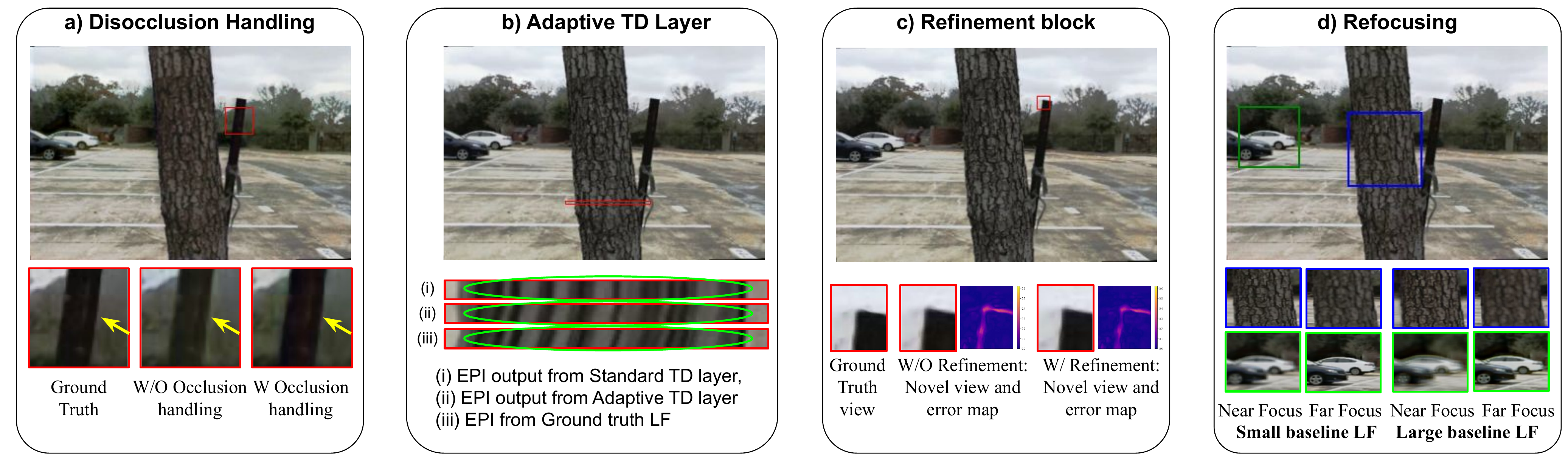}
    \caption{
    We propose three novel techniques: a) disocclusion handling, b) adaptive low-rank representation for \gls{lf} and c) a novel refinement block, for \gls{lf} video reconstruction from monocular video.
    Combining these with the self-supervised cost functions inspired by \cite{shedligeri2021selfvi}, we can reconstruct high-fidelity \gls{lf} videos, even with varying baselines.
    As shown in (d) this allows us to control the synthetic defocus blur using the output video.
    }
    \label{fig:fig-1}
\end{figure}

\section{Introduction}
Cameras have become cheap and ubiquitous in the modern world, giving consumers a capability to acquire photos and videos anywhere and anytime.
The last decade saw an accelerated improvement in image sensors and lens quality, leading to a significant improvement in the picture quality from these tiny cameras.
Towards the end of the decade, the focus shifted towards more and more innovative software, pushing the limits to what can be achieved with these ubiquitous cameras \cite{delbracio2021mobile}.
This push resulted in a variety of features: ranging from simple effects like background-blur to more dramatic ones like augmented reality.
Features like bokeh effects and novel view synthesis became popular as they provided a sense of `$3$D' to the otherwise flat pictures.
However, these features have currently been limited to images and there's no straightforward way of extending them to videos.
In the last few years, videos have certainly become a more powerful means of communication, knowledge-sharing and even entertainment.
\gls{lf} imaging could provide an intuitive way of bringing these features to videos.
However, there's no easy way to capture \gls{lf} videos yet.
Computational photography is poised to solve this, making it easy and accessible to capture \gls{lf} on small form-factor devices \cite{kim2020miniaturized}.
We instead focus on existing camera hardware and aim to reconstruct \gls{lf} videos from any ordinary monocular camera.

Traditionally, \gls{lf} imaging required use of bulky or complex hardware setups such as camera arrays \cite{wilburn2005high} and micro-lens arrays \cite{ng2005light}.
Hence, the recent focus has been on reducing the hardware complexity through the use of learning-based techniques.
Typically, these involve the reconstruction of \gls{lf} from sparse input views (such as 1, 2 or 4 views) \cite{kalantari2016learning,zhang2015light,srinivasan2017learning,li2020synthesizing}.
To solve the challenges in acquiring \gls{lf} videos through commercial cameras, several techniques for \gls{lf} \emph{video} reconstruction have also been proposed \cite{bae20215d,wang2017light,shedligeri2021selfvi}.
\stereo\ \cite{shedligeri2021selfvi} is an interesting recent work that proposes a novel self-supervised technique for \gls{lf} video reconstruction from stereo videos.
Being a self-supervised technique it relied on an intermediate low-rank representation for \gls{lf} frames achieving high-quality reconstructions.
However, it requires a stereo video input where both cameras should have identical focal lengths (identical field-of-view).
This can become a limitation considering that stereo cameras are still not as widespread as monocular cameras.
This is especially true for consumer applications, where mostly monocular cameras are preferred. 

Motivated by the availability of large and diverse sets of high-quality monocular videos we propose a novel, self-supervised learning technique for \gls{lf} video reconstruction from monocular input.
To start with, we preserve the self-supervised photometric, geometric and temporal consistency constraints adopted in \stereo\ (see \cref{sec:loss_fns}).
Further we introduce three crucial blocks that are necessary for our case of monocular video input. These are: 1) a novel loss for handling disocclusion, 2) a scene geometry adaptive intermediate low-rank representation and 3) a novel and supervised refinement block to further refine the LF video(see \cref{fig:fig-1}).

The challenge with just a monocular input is that there's no information on how to fill the disoccluded regions/pixels of the predicted \gls{lf}.
We propose a technique to \emph{inpaint} the disoccluded regions of the estimated \gls{lf} frames.
The intuition is that, in a video acquired using a moving camera, occluded regions in one frame might be visible in the neighboring temporal frames.
Our disocclusion handling technique (\cref{sec:disocclusion}) utilizes this existing information to fill in the disoccluded regions of the \gls{lf} frame. 

Next, we modify the standard \gls{td} based intermediate low-rank representation so that it can adapt to any input scene.
While \gls{td} model \cite{wetzstein2012tensor} uses fixed displacement between the layers, we propose a modification where this displacement can be modified for each input image (\cref{sec:adaptive_td}).
In \cite{wetzstein2012tensor}, each of the layers are shown to represent a depth-plane in the scene.
Hence, by estimating the displacement values for each scene, the layers are better able to represent the given \gls{lf}.
This idea was inspired from a similar choice of adaptive layered representation in \cite{li2020synthesizing} for novel view synthesis.
Unlike \cite{li2020synthesizing} we adopt a more sophisticated approach to predict the depth planes through global scene understanding by using transformers \cite{bhat2021adabins}.
As shown in our experiments, the adaptive low-rank representation provides a significant boost in the quality of the predicted \gls{lf} frames.

Finally, we explore the popular idea of self-supervised pre-training, followed by supervised learning on a small amount of data to boost the performance of a model \cite{chen2020simple}\cite{caron2020unsupervised}\cite{jaiswal2021survey}.
We design a novel convolutional vision-transformer-based \cite{dosovitskiy2020image} refinement block that is trained via supervised learning on a small amount of \gls{lf} \emph{image} data.
This helps in further refining the output around the depth-edges that are difficult to reconstruct with just self-supervised learning.
The final output is a weighted combination of the refinement block output and the \gls{lf} estimated by self-supervised learning (\cref{sec:refinement}).
To the best of our knowledge, this is the first time that vision transformers are used to supervise \gls{lf} reconstruction by efficiently combining the spatio-angular information.
In summary, we make the following contributions:
\begin{itemize}
    \item[\textbullet] High quality reconstruction of \gls{lf} videos from monocular video with self-supervised learning.
    \item[\textbullet] Handling disocclusions in rendering \glspl{lf} using self-supervised consistency losses utilizing information from successive video frames.
    \item[\textbullet] A modified \gls{td}-based low-rank representation that can adapt to the given input scene dynamically adjusting the distance between the layers.
    \item[\textbullet] A novel supervised vision-transformer based refinement block to exploit the small amount of \gls{lf} image data to further improve reconstruction on video.
\end{itemize}

\section{Related Work}
\label{sec:related}
\paragraph{\textbf{\gls{lf} synthesis}}
While the concept of \gls{lf} or integral imaging is quite old  \cite{lippmann:jpa-00241406,adelson1991plenoptic}, capturing these images has been complicated.
While commercial \gls{lf} cameras are now available in the market \cite{ng2005light}, they suffer from low spatial resolution.
Over the last several years, a diverse set of camera setups and algorithms have aimed at making \gls{lf} imaging simpler and more accessible.
There have been setups that use coded-aperture systems \cite{veeraraghavan2007dappled,inagaki2018learning,sakai2020acquiring}, cameras with coded masks near the sensor \cite{marwah2013compressive,hajisharif2020single} and even hybrid sensors \cite{wang2017light}.
Later, with advances in deep-learning, systems using ordinary commercial cameras such as one or more DSLRs became popular.
Techniques that reconstruct \gls{lf} frames from focus-defocus pair \cite{vadathya2019unified} or focal-stack images\cite{blocker2018low} were proposed.
Several techniques were also proposed that could reconstruct \gls{lf} from sparse set of views on a regular grid. The number of views could be 1-view\cite{srinivasan2017learning,li2020synthesizing,bae20215d,ivan2019synthesizing}, 2-views\cite{zhang2015light,shedligeri2021selfvi}, 4-views\cite{kalantari2016learning,Wang_2018_ECCV,Yeung_2018_ECCV} and even 9-views\cite{wu2017light}.

\paragraph{\textbf{\gls{lf} synthesis from monocular image}}
As ordinary monocular cameras are ubiquitous, several techniques aim at \gls{lf} reconstruction from them.
As this is an ill-posed problem, learning-based techniques have been essential in this domain.
A popular technique has been to first predict disparity flow \cite{srinivasan2017learning} or appearance flow\cite{ivan2019synthesizing}\cite{zhou2016view} and then warp the input image accordingly to reconstruct the \gls{lf} frame.
Recently, \gls{mpi} based representation is being used for \gls{lf} prediction \cite{huang2018deepmvs,zhou2018stereo,srinivasan2019pushing,mildenhall2019local,li2020synthesizing}.
Li \textit{et al.} \cite{li2020synthesizing} propose a modified \gls{mpi} model that allowed them to significantly reduce the representation complexity.
With a similar intuition, we propose a modified low-rank representation based on layered \gls{lf} displays \cite{wetzstein2012tensor} for predicting the \gls{lf} frames.

\paragraph{\textbf{\gls{lf} video reconstruction}}
As commercial \gls{lf} cameras such as Lytro acquire videos at only $3$ \gls{fps}, \gls{lf} video acquisition at high angular and temporal resolution has also been challenging.
In \cite{wang2017light} a learning-based algorithm with a hybrid camera system consisting of a general DSLR camera and a light field camera was proposed.
Hajisharif \textit{et al.} \cite{hajisharif2020single} proposed a single sensor-based algorithm that required a coded mask to be placed in front of the sensor.
As these algorithms require complex and bulky hardware setups, techniques such as \cite{bae20215d,shedligeri2021selfvi} are proposed that just require ordinary cameras.
Although our self-supervised algorithm is inspired from \cite{shedligeri2021selfvi}, the closest work to ours is \cite{bae20215d}.
Bae \textit{et al.}\cite{bae20215d} utilize a large set of computer-generated data to supervise a neural network for \gls{lf} video reconstruction from monocular video.
In contrast, our proposed technique does not require hard-to-acquire \gls{lf} video data for supervision.

\paragraph{\textbf{Learning with layered LF representation}}
Previously, layered LF display representations \cite{wetzstein2012tensor} have been used in conjunction with neural networks. 
\cite{maruyuma2019a} built an end-to-end pipeline from a coded aperture scene acquisition for displaying the scene on a layered LF display. 
Similar work in \cite{takahashi2018from,kobayashi2017from} aims at capturing a focal stack and then learning to display the scene onto the LF display.
Inspired by \cite{shedligeri2021selfvi}, we also adopt the layered LF display based intermediate low-rank representation \lowrank\ for \gls{lf} estimation.
We extend the standard low-rank model to adapt to the individual scene by predicting the optimal distance between the layers for each input image.

\section{Monocular LF video estimation}
\label{sec:method}

\begin{figure}[t!]
    \centering
    \includegraphics[width=\textwidth]{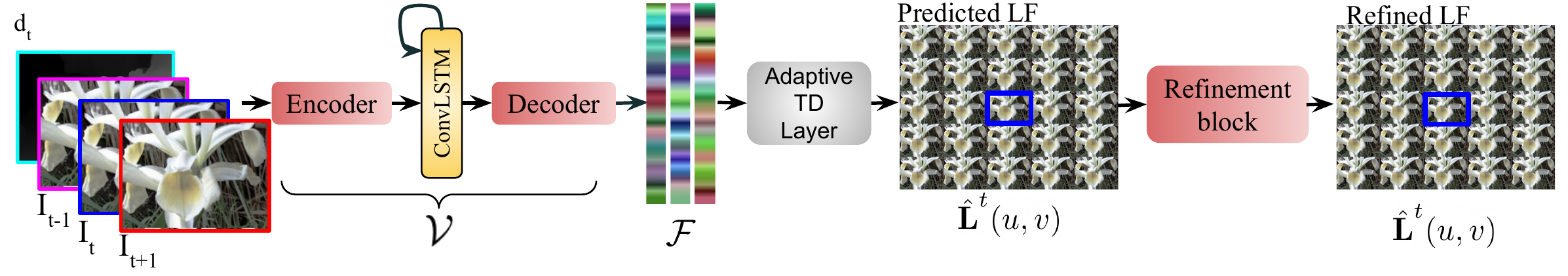}
    \caption{Our proposed algorithm takes as input a sequence  $\mathcal{I} = $ \{\prevframe, \currframe, \nextframe, \currdisp\}.
    A recurrent \gls{lf} synthesis network \lfnet\ first predicts an intermediate low-rank representation \lowrank\ for the corresponding \gls{lf} frame. 
    An adaptive TD layer (\ref{sec:adaptive_td}) takes the same set $\mathcal{I}$ and \lowrank\ as input and outputs the \gls{lf} frame \estcurrlf.
    A set of self-supervised cost-functions (\ref{sec:loss_fns}, \ref{sec:disocclusion}) are then imposed on \estcurrlf\ for the end-to-end training of \lfnet and the adaptive TD layer.
    Finally, a refinement block (\ref{sec:refinement}) then takes \estcurrlf\ and \currframe\ as input and outputs a refined \gls{lf}.
    }
    \label{fig:model}
\end{figure}

\begin{figure}[t!]
    \centering
    \includegraphics[width=\columnwidth]{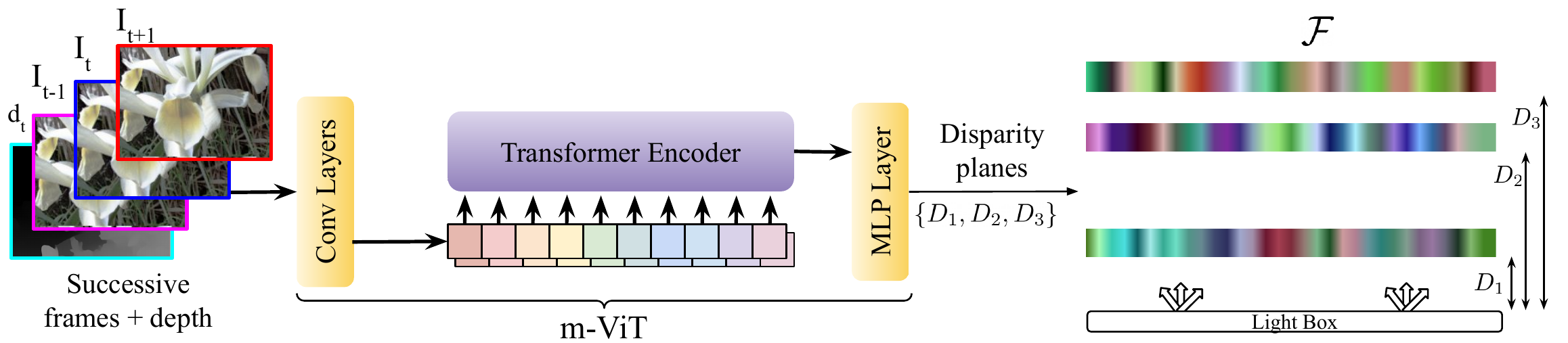}
    \caption{We use an \gls{mVIT} block \cite{bhat2021adabins} to predict the displacement between the different layers of the \gls{td} based low-rank representation. 
    Each of the layers in \lowrank\ approximately represent a scene depth plane. Instead of keeping the layers static/fixed, a scene-specific displacement value will move the layer to a depth plane where it can best represent the scene.
    }
    \label{fig:adaptive_td_layer}
\end{figure}

We propose a self-supervised learning based technique for \gls{lf} video reconstruction from a monocular video sequence.
For each input frame of the monocular video, we reconstruct a corresponding \gls{lf} video frame.
As shown in \cref{fig:model}, a deep neural network takes as input, a sequence of $3$ input frames and a disparity map \{\prevframe, \currframe, \nextframe, \currdisp\} and estimates an intermediate low-rank representation of the current \gls{lf} frame \estcurrlf.
As shown in \cref{fig:adaptive_td_layer} and further elaborated in \cref{sec:adaptive_td}, we propose a modified intermediate low-rank representation adapted from \cite{wetzstein2012tensor}.
After obtaining \estcurrlf\ from the adaptive \gls{td} layer, we introduce the geometric, photometric and the temporal consistency constraints \cite{shedligeri2021selfvi} to train our \gls{lf} synthesis network (see \cref{sec:loss_fns}).
Being a self-supervised technique, we do not have any information about the disoccluded regions in {\estcurrlf} from just {\currframe}. 
Hence, we introduce a disocclusion handling technique that utilizes information from {\prevframe} and {\nextframe} to fill-in the disoccluded regions of \estcurrlf (see \cref{sec:disocclusion}).
Finally, to further refine the estimated \gls{lf} frame, we propose a novel residual refinement block based on vision-transformers (see \cref{sec:refinement}) which is trained using supervised learning.


\subsection{Light field frame prediction}
\label{sec:lfnet}
As shown in \cref{fig:model}, we stack three successive input frames and the corresponding disparity map as $\mathcal{I}=$\{\prevframe, \currframe, \nextframe,\currdisp\} and feed it to the \gls{lf} prediction network \lfnet.
With a monocular input, it's not possible to obtain a disparity map directly.
Hence, we first estimate a relative depth map {\currdepth} of {\currframe} using a pre-trained monocular depth estimation model \cite{ranftl2021vision}.
We know that a disparity map is related to the depth map up to an affine transformation \cite{garg2019learning}, defined here as $\currdisp = a \currdepth + b$, where $a$ and $b$ are two scalars.
During training, we randomly sample values of $a$ and $b$ and convert the relative depth map $\currdepth$ to the disparity map $\currdisp$.
The network \lfnet is a \gls{lstm} based model consisting of an encoder and decoder with skip connections.
The network \lfnet predicts an intermediate low-rank representation \lowrank\ for \estcurrlf\ based on a modified tensor-display model \cite{wetzstein2012tensor}.
We describe the process of obtaining \estcurrlf\ from the low-rank representation \lowrank\ in \cref{sec:adaptive_td}.

\subsection{Adaptive tensor-display model}
\label{sec:adaptive_td}
In the previous section, we estimated the representation \lowrank\ from the network \lfnet, based on the low-rank model proposed in \cite{wetzstein2012tensor}.
In this standard model,  $\lowrank = [f_{-N/2}, \ldots, f_0, \ldots, f_{N/2}]$, where $f_k = [f_k^1, f_k^2, \ldots, f_k^R]$, $f_k^r \in [0,1]^{h\times w\times 3}$.
Here $N$ represents the number of layers in the low-rank model and $R$ represents its corresponding rank.
Given \lowrank, the corresponding \ford \gls{lf} frame can be computed as
\begin{equation}
    L(x,y,u,v) = \td(\lowrank) = \sum_{r=1}^R\prod_{n=-{N/2}}^{{N/2}} f_{n}^r(x+nu,y+nv)
\end{equation}
where ${x,y}$ and ${u,v}$ respectively denote the spatial and angular coordinates.
Further analysis into these representations in \cite{wetzstein2012tensor} showed that each layer approximately represents a particular depth plane in the scene.
However, the standard model places these layers at a uniform distance from each other representing depth planes placed uniformly in the scene.
In a natural image the objects in the scene could be distributed non-uniformly throughout the depth.
This idea was exploited in \cite{li2020synthesizing}, where the standard \gls{mpi} model was adapted to each input image by assigning non-uniform disparity values for each \gls{mpi} layer.
This drastically reduced the number of \gls{mpi} layers required to represent the scene up to a similar accuracy.

Motivated by this we use a \gls{mVIT} network \cite{bhat2021adabins}, to predict a sequence of values, $D=\{D_{-N/2}, \ldots D_{N/2}\}$, that will be used in adapting the \td layer to each input (\cref{fig:adaptive_td_layer}).
\gls{mVIT} predicts one value for each layer in the representation \lowrank\ using the input $\mathcal{I}=\{\prevframe, \currframe, \nextframe, \currdisp\}$.
The values in $D$ are used in the proposed adaptive \td layer as
\begin{equation}
    L(x,y,u,v) = \td(\lowrank; D) = \sum_{r=1}^R\prod_{n=-{N/2}}^{{N/2}} f_{n}^r(x+D_nu,y+D_nv) ~,
\end{equation}
where $D_n$ represents the scalar value predicted by \gls{mVIT} for layer $n$.
After computing \estcurrlf\ from our proposed adaptive \td layer, we impose three main self-supervised cost functions to train the prediction network \lfnet.

\subsection{Loss functions}
\label{sec:loss_fns}
To successfully train the \gls{lf} prediction network \lfnet, we follow \cite{shedligeri2021selfvi} and define three constraints that enforce the structure of the \gls{lf} video on the predicted sequence of frames.

\subsubsection{Photometric constraint}
The photometric constraint is defined on the premise that the center view of {\estcurrlf} should match the current input frame \currframe.
Hence, we define the loss function reflecting this as $\loss_{photo}^t = \| \estcurrlf \left(\mathbf{0}\right) - \currframe\|_1$, where $\estcurrlf\left(\mathbf{0}\right)$ represents the central angular view of \estcurrlf.

\subsubsection{Geometric constraint}
To compute the cost, we first warp all \glspl{sai} of the {\estcurrlf} to the \gls{sai} $\mathbf{0}$ that corresponds to \currframe.
In essence, we warp $\estcurrlf(\mathbf{u})$ to the \gls{sai} $\mathbf{0}$ to obtain $\estcurrlf(\mathbf{u}\shortrightarrow \mathbf{0})$, expressed as,
\begin{equation}
{
    \estcurrlf(\mathbf{u}\shortrightarrow \mathbf{0}) = \warp\left(\estcurrlf\left(\mathbf{u}\right); \left(\mathbf{u}-\mathbf{0}\right)\currdisp \right) ~.
}
\end{equation}
Here, {\warp} denotes the bilinear inverse warping operator~\cite{jaderberg2015spatial} that takes as input a displacement map and remaps the images.
The geometric consistency error between the approximated current frame $\estcurrlf(\mathbf{u}\shortrightarrow \mathbf{0})$ and {\currframe} is then defined as,
\begin{equation}
    \loss_{geo}^t = \sum_{\mathbf{u}} \| \estcurrlf\left(\mathbf{u}\shortrightarrow \mathbf{0}\right) - \currframe \|_1  ~.
    \label{eq:disp-loss}
\end{equation}

\subsubsection{Temporal consistency constraint}
In addition to an \gls{lstm} network-based \cite{shi2015convolutional} recurrent framework of our network \lfnet, we impose a temporal consistency constraint on the predicted outputs.
For this, we first estimate the optical flow between successive input video frames using a pre-trained RAFT \cite{teed2020raft} network, denoted as \flownet.
The optical flow is then computed as $o_t = \flownet(\currframe, \nextframe)$.
To enforce temporal consistency, we utilize the warped angular views $\estcurrlf\left(\mathbf{u}\shortrightarrow \mathbf{0}\right)$ and
again warp them to the video frame at $t+1$ using $o_t$.
Then, the temporal consistency error is defined as the error between the known next frame {\nextframe} and these warped frames and is denoted as,
\begin{equation}
    \loss_{temp}^t = \sum_{\mathbf{u}} 
 \| \warp\left(\estcurrlf\left(\mathbf{u} \shortrightarrow \mathbf{0} ; o_t\right) \right) - \nextframe \|_1 ~.
    \label{eq:temp-loss}
\end{equation}
Minimizing this error during training explicitly enforces temporal consistency between the successive predicted \gls{lf} frames.

\begin{figure}[t]
    \centering
    \includegraphics[width=\columnwidth]{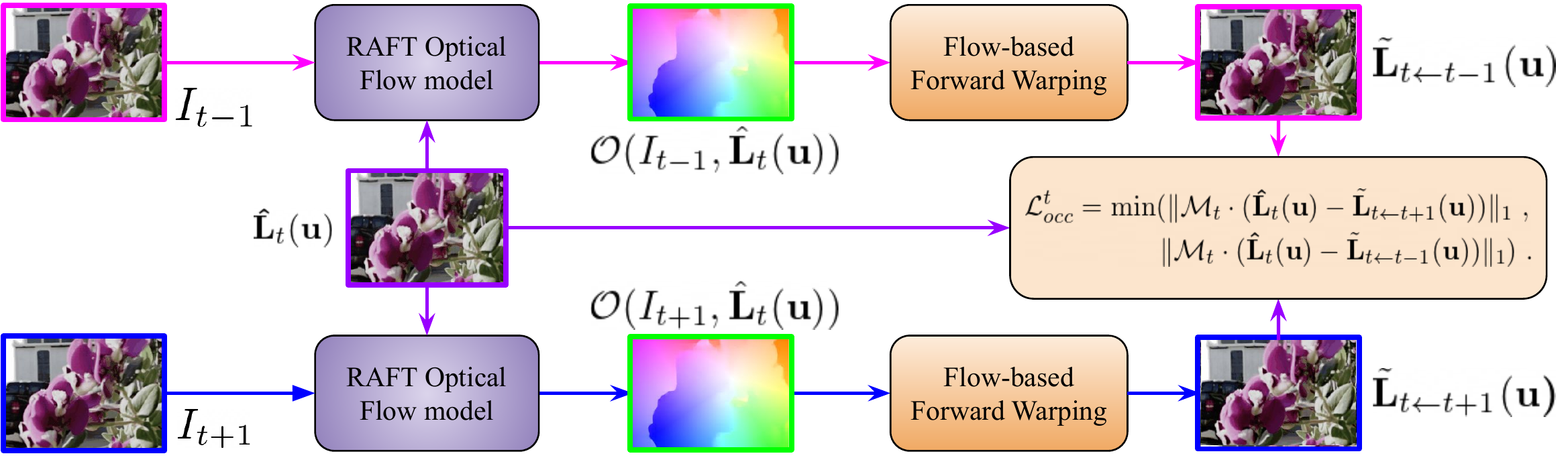}
    \caption{We introduce disocclusion handling to constrain the synthesis network to fill in the disoccluded pixels of the \gls{lf} with information from the neighboring frames.
    As shown, \prevframe\ is forward warped to each view of \estcurrlf, using the optical flow predicted from RAFT \cite{teed2020raft}.
    We also warp \nextframe\ to each \glspl{sai} of \estcurrlf, and the loss is computed as in \cref{eq:min_reprojection}.
    }
    \label{fig:occlusion}
\end{figure}

\subsection{Disocclusion handling}
\label{sec:disocclusion}

In a \gls{lf}, pixels at depth boundary of objects get occluded and disoccluded between different \glspl{sai}.
Due to the lack of ground truth data we face a major challenge when learning to fill-in the intensity values at the disoccluded pixels.
Our objective is to use pixels in \prevframe\ and \nextframe\ to fill in the disoccluded pixels of \estcurrlf, as these frames could potentially have the necessary pixel values.
Pixels from neighboring video frames have been used to inpaint the current frame in several video-inpainting techniques \cite{kim2019deep,xu2019deep}.
Here, we achieve this by bringing the informative intensity values to the disoccluded pixels through flow-based warping.
As shown in \cref{fig:occlusion}, we use RAFT to obtain optical flow between the \glspl{sai} of \estcurrlf\ and the input frames \prevframe, \nextframe.
By including these warping based operations in a loss function, we train the network to automatically predict the disoccluded pixels.

The loss is defined on only those pixels that are disoccluded in the \glspl{sai} of the \gls{lf} frame \estcurrlf.
We obtain the disoccluded pixels by forward-warping the input frame \currframe\ to all the \glspl{sai} of \gls{lf} with the disparity map \currdisp.
For each \gls{sai} at angular location $\mathbf{u}$, we define a binary mask \currmask\ which is $1$ if forward warping resulted in a hole for that particular pixel.
To fill the dis-occluded pixels, we forward warp \prevframe to the predicted \glspl{sai} of \estcurrlf\ using optical flow as shown in \cref{fig:occlusion}.
The forward warped \glspl{sai} are obtained as:
\begin{equation}
    \frwdwarpprevframelf = \warp \left(\prevframe; \flownet \left(\prevframe, \estcurrlf(\bfu) \right) \right),
\end{equation}
\begin{equation}
    \frwdwarpnextframelf = \warp \left(\nextframe; \flownet \left(\nextframe, \estcurrlf(\bfu) \right) \right),
\end{equation}
where \frwdwarpnextframelf\ and \frwdwarpprevframelf\ represent the forwards warped \gls{sai}s from \nextframe\ and \prevframe\ respectively using optical flow.
Depending on the camera motion the disoccluded pixels in \estcurrlf could be visible either in \prevframe\ or \nextframe\ or both.
Taking this into consideration, we define the cost function as
\begin{equation}
        \loss_{occ}^{t} = \min(\|\currmask \cdot \left(\estcurrlf(\bfu) - \frwdwarpprevframelf\right)\|_1 ~, ~ \|\currmask \cdot \left(\estcurrlf(\bfu) - \frwdwarpnextframelf\right)\|_1 )
        \label{eq:min_reprojection}
\end{equation}
$\loss_{occ}^{t}$ follows the concept of minimum re-projection loss followed in monocular depth estimation techniques such as \cite{godard2019digging}.

\begin{figure}[t]
    \centering
    \includegraphics[width=\columnwidth]{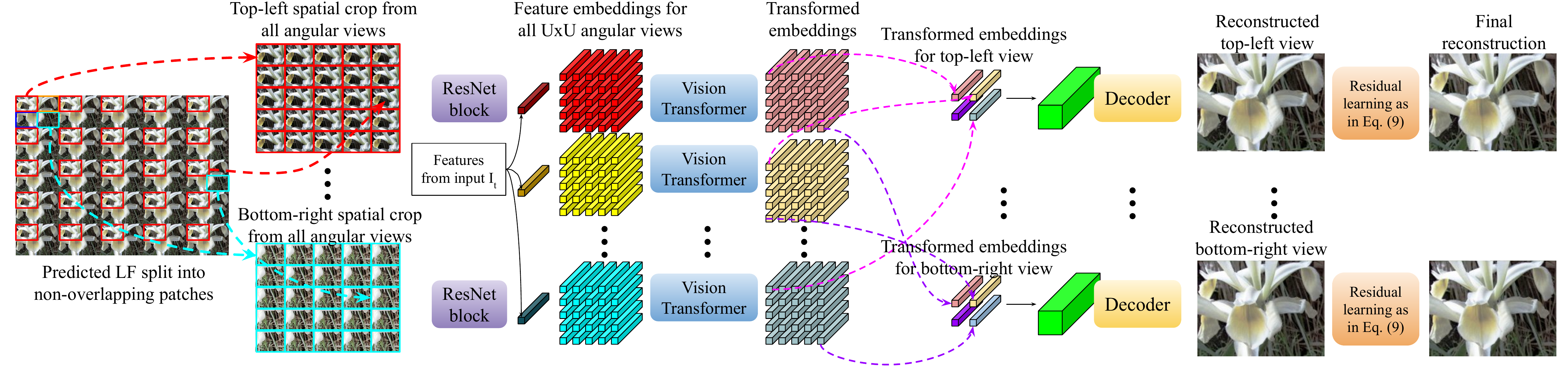}
    \caption{
    A supervised residual refinement block is used to further improve the reconstruction quality of the \glspl{lf}.
    The transformer block attends to the spatio-angular information in the estimated \gls{lf} and the input frame \currframe\ to predict the refined output.
    }
    \label{fig:refinement}
\end{figure}

\subsection{Supervised residual refinement block}
\label{sec:refinement}

Recently, self-supervised pre-training on very large unlabeled datasets followed by supervised learning on a limited labeled dataset has helped in achieving state-of-the-art results \cite{chen2020simple,caron2020unsupervised,jaiswal2021survey}.
Inspired by these works, we propose to use the limited dataset of \gls{lf} \emph{images} to further refine the reconstructed \gls{lf} frames.
As this shouldn't affect the temporal consistency of the predicted frames, the proposed refinement module follows a residual network architecture as shown in \cref{fig:refinement}.
And this module can be trained as a separate block from the recurrent module in the synthesis network \lfnet.

Vision Transformers(ViT) \cite{dosovitskiy2020image} form the backbone of our proposed refinement module.
As shown in \cref{fig:refinement}, we divide the predicted \gls{lf} frame \estcurrlf\ into non-overlapping patches, each of size $p\times p$.
For simplicity consider all the $U^2$ top-left patches cropped from each angular view of {\estcurrlf}.
A shallow ResNet-based neural network extracts features independently from each of the $U^2$ patches.
Additionally, we also extract features from the top-left patch of the input image {\currframe}.
The transformer module then takes as input the $U^2+1$ features/embeddings as input and outputs $U^2+1$ tokens after applying \gls{mhsa} \cite{dosovitskiy2020image}.
An identical procedure is repeated on all the non-overlapping patches of {\estcurrlf} to produce $U^2+1$ tokens each time.

As in \cref{fig:refinement}, we discard the token from the input frame and consider all the $P$ transformed tokens from a particular angular view, say bottom-right.
Here, $P$ is the number of non-overlapping patches cropped from each angular view.
These $P$ tokens are stacked horizontally and vertically following the order of cropped patches, so as to form a larger feature map.
A shallow decoder network then takes these stacked tokens as input and predicts a $4$ channel output.
The first $3$ channels form an RGB image ($ \outputref(\mathbf{u})$) and the fourth channel is the mask $M_{ref}\in [0,1]^{h\times w}$.
The final output {\estcurrlfref} is then defined as,
\begin{equation}
    \estcurrlfref(\mathbf{u}) = M_{ref}\odot \estcurrlf(\mathbf{u}) + (1 - M_{ref}) \odot \outputref(\mathbf{u}) ~.
\end{equation}
Identical decoding step is repeated for each \gls{sai} $\mathbf{u}$ producing a refined \gls{lf} frame \estcurrlfref.
As we assume access to a \gls{lf} \emph{image} dataset, we train the refinement network by imposing L1 loss between \estcurrlfref\ and the corresponding ground-truth \truelf\ as:
\begin{equation}
    \loss_{ref} = \sum_{\mathbf{u}} \| \estcurrlfref(\mathbf{u}) - \truelf(\mathbf{u}) \|_1 ~.
    \label{eq:refinement}
\end{equation}

\subsection{Overall loss}
We finally add total-variation(TV)-based smoothness constraint\cite{shedligeri2021selfvi} on the predicted LF frames and Bin-center density loss \cite{bhat2021adabins} on disparity values predicted by \gls{mVIT}.
The Bin-center density loss encourages the predicted disparity planes to be close to the disparity map $d_t$ which is provided as input to the adaptive \td\ layer.
Including all the cost functions, the overall loss to minimize for training \lfnet\ and the adaptive \gls{td} layer becomes,
\begin{equation}
    \loss_{self}^t =  \lambda_1 \loss_{photo}^t + \lambda_2 \loss_{geo}^t + \lambda_3 \loss_{temp}^t + \lambda_4 \loss_{occ}^t + \lambda_5 \loss_{bins}^t + \lambda_6 \loss_{TV}^t~,
    \label{eq:overall_loss}
\end{equation}
where the parameters $\lambda_i$ control the contribution of each loss term.
After the self-supervised training of the main network is completed, we then freeze these weights and train the refinement block.
The refinement block is trained using a supervised cost function $\loss_{ref}$ in \cref{eq:refinement}.

\subsection{Implementation details}
\label{sec:implementation}
As shown in \cref{fig:model}, our proposed pipeline has three separate deep neural networks: (a) \gls{lf} synthesis network, (b) adaptive \td\ layer (\cref{fig:adaptive_td_layer}) and (c) refinement network (\cref{fig:refinement}).
The synthesis network \lfnet\ is a \gls{lstm} based recurrent neural network consisting of a Efficient-Net encoder \cite{tan2019efficientnet} and a convolutional decoder with skip connections.
In the adaptive \gls{td} layer, we set the low-rank representation {\lowrank} to have $N=3$ layers and the rank $R=12$ following \cite{shedligeri2021selfvi}.
The displacements $D=\{D_1, D_2, D_3\}$ are predicted from \gls{mVIT}\cite{bhat2021adabins} network that takes as input \{\prevframe,\currframe,\nextframe,\currdisp\}.
Finally, the refinement network has a backbone of the convolutional vision transformer which is supervised using a limited amount of \gls{lf} \emph{image} data.
Further details of the neural networks can be found in the supplementary material.

For training our proposed synthesis network, we use the \emph{GOPRO} monocular video dataset\cite{Nah_2017_CVPR}.
The \emph{GOPRO} dataset contains monocular videos of 33 different scenes each containing 525 to 1650 monocular frames of spatial resolution $720 \times 1280$.
We split the dataset into a set of 25 videos for training and 8 videos for validation.
The monocular video frames are resized into frames of size $352 \times 528$ to maintain the spatial resolution of Lytro Illum light field camera.
While training we obtain a monocular video of 7 frames and randomly crop a patch of size $176 \times 264$.
The successive frames in the training data are $10$ frames apart in the raw GoPro videos captured at $240$ \gls{fps}.
This ensures that there's reasonable object motion between successive input frames which is crucial for the disocclusion handling technique.
In one frame, closer objects show larger disocclusions in the predicted \gls{lf} as they have higher disparity values.
These objects also proportionally have larger displacements in successive frames, providing enough information to fill in the disoccluded pixels.

The relative depth map input to the network is obtained from \cite{ranftl2021vision} and then modified for various baseline factors to enable the synthesis network to generate \gls{lf} outputs of various baseline.
We randomly choose a value for $a\in\{0.8, 1.6, 2.4, 3.2\}$ and $b\in[0.2,0.4]$ to obtain disparity $d_t=az_t + b$ as explained in \cref{sec:lfnet}.
The network is trained in Pytorch\cite{Pytorch} using AdamW \cite{loshchilov2017decoupled} optimizer for 25 epochs, with an initial learning rate of $0.0001$ and weight decay of $0.001$. 
The learning rate is decreased to half the initial value when the validation loss plateaus for more more than 4 epochs.
We empirically choose the hyperparameters as $\lambda_1 = 1.0$, $\lambda_2 = 1.0$, $\lambda_3 = 0.5$, $\lambda_4 = 0.2$, $\lambda_5 = 2$ and $\lambda_6 = 0.1$ in \cref{eq:overall_loss}.

For training our residual refinement block, we freeze the weights of the synthesis network and train only the refinement block using supervised loss function in \cref{eq:refinement}.
We fix the value of $a$ as $1.2$ and $b$ as $0.3$ to estimate $d_t$ which is provided as input to the synthesis network.
For the supervised training, we use $1000$ \gls{lf} images from \emph{TAMULF} \cite{li2020synthesizing} dataset.
The network is trained using AdamW optimizer for 15 epochs, with an initial learning rate of $0.001$ and weight decay of $0.001$. 
The learning rate is decreased to half the initial value when the validation loss plateaus for more more than $4$ epochs.

\section{Experiments}
\label{sec:expts}
To validate our proposed algorithm, we make several qualitative and quantitative comparisons with diverse \gls{lf} datasets.
For quantitative comparison, we mainly consider four different datasets: \emph{Hybrid} \cite{wang2017light}, \emph{ViewSynth} \cite{kalantari2016learning}, \emph{TAMULF} \cite{li2020synthesizing} and \emph{Stanford} \cite{dansereau2019liff} containing 30, 25, 84 and 113 light field video sequences, respectively.
From the \emph{Hybrid} dataset we consider the central $7\times 7$ views as the ground-truth light field videos, and the center-view of each \gls{lf} forms the input monocular video.
The rest three datasets are \gls{lf} \emph{image} datasets, and we simulate \gls{lf} videos with $8$ frames from each \gls{lf} following the procedure described in \cite{shedligeri2021selfvi}\cite{lumentut2019deep}.
The center-view of these $7\times 7$ view \gls{lf} videos form the monocular video sequence that is given as input to our algorithm.
During inference, we first obtain the depth estimate \currdepth\ from DPT\cite{ranftl2021vision} and convert it to a disparity map \currdisp.
Three consecutive temporal frames and disparity map are stacked and input to the complete model represented in \cref{fig:model} to obtain the \gls{lf} frame output.

\subsection{Light field video reconstruction}
\label{sec:quant}
We quantitatively and qualitatively compare the results of our proposed algorithm with previously proposed monocular \gls{lf} estimation techniques.
For quantitative comparison, we use two metrics: peak signal-to-noise ratio (PSNR) (higher is better) and structural similarity index measure (SSIM) (higher is better).
As shown in \cref{tab:quant}, we compare the performance of our proposed algorithm with Niklaus \textit{et al.} \cite{niklaus20193d}, Srinivasan \textit{et al.} \cite{srinivasan2017learning} and Li \textit{et al.} \cite{li2020synthesizing}.

Li \textit{et al.} \cite{li2020synthesizing} takes a single frame and a relative depth estimate from \cite{DBLP:journals/corr/abs-1810-08100} as input.
To obtain the complete \gls{lf} video, we have to reconstruct each frame of the video individually.
In \cref{tab:quant}, Li \textit{et al.} + Ranftl \textit{et al.} represents a modified \cite{li2020synthesizing}, where we input a depth estimate from DPT\cite{ranftl2021vision} instead of the original DeepLens\cite{DBLP:journals/corr/abs-1810-08100} model.
This is done to ensure a fair comparison with our technique as we also use DPT, which is a state-of-the-art monocular depth estimation technique based on vision transformers.
However, \cite{li2020synthesizing} is not trained for inputs from DPT \cite{ranftl2021vision}.
Hence, we finetune \cite{li2020synthesizing} on the TAMULF dataset with depth maps from DPT \cite{ranftl2021vision}.
Srinivasan \textit{et al.} \cite{srinivasan2017learning} is another single image \gls{lf} estimation model. 
While the original network is trained on a dataset of flower images (proposed in the same work), we finetune it on a larger and diverse TAMULF dataset from \cite{li2020synthesizing}.
Finally, we also compare our algorithm with Niklaus \textit{et al.} \cite{niklaus20193d} that takes a single frame as input.
We used the default implementation provided by the authors for comparison, which is already trained on a diverse dataset.
As all these techniques are image-based and don't have any temporal information, we also compare with a \emph{downgraded} version of our algorithm `Proposed(image)'.
In this model, we repeat the current frame as three successive input frames of our proposed algorithm.

\begin{figure}[t]
    \centering
    \includegraphics[width=\columnwidth]{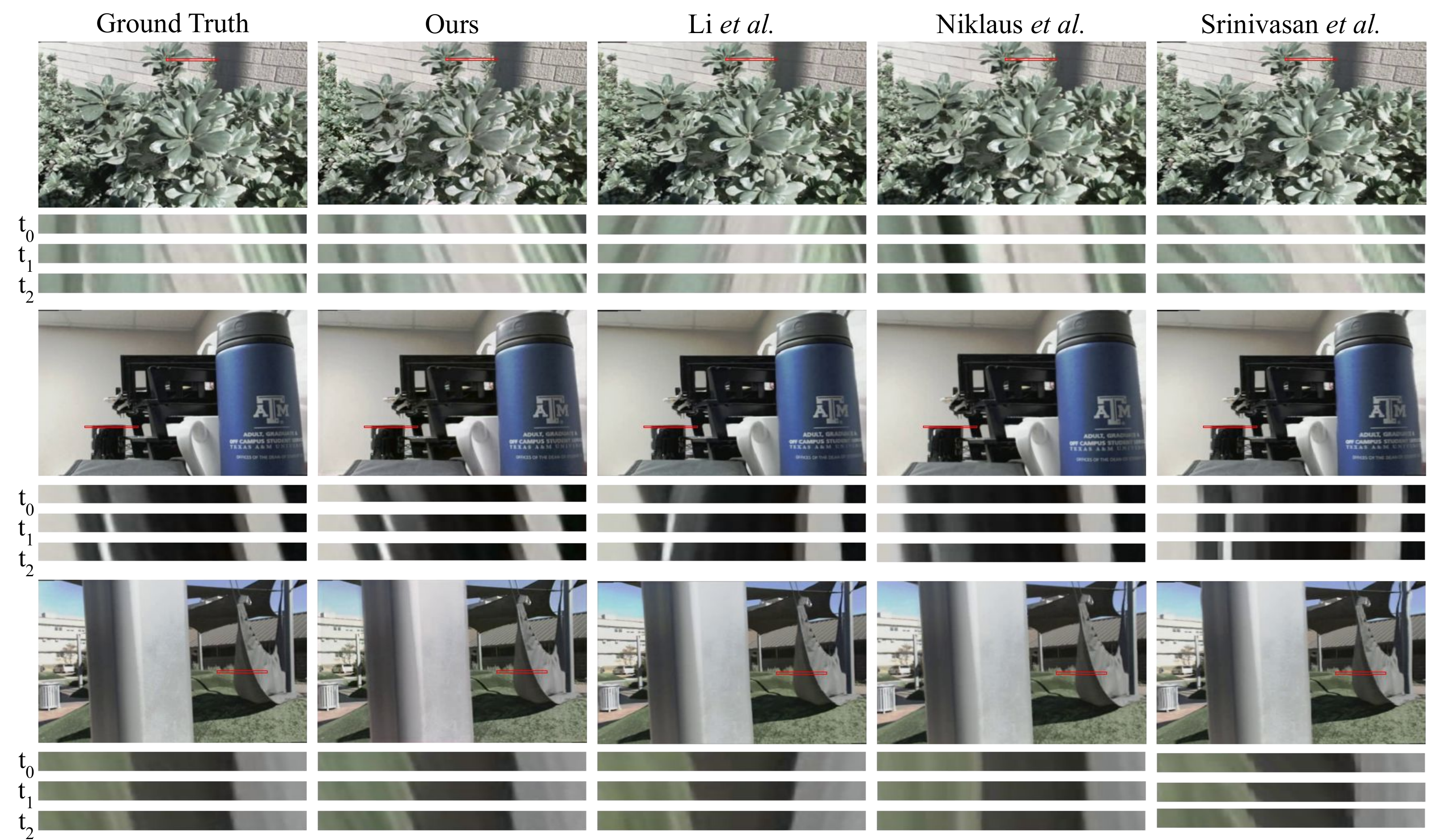}
    \caption{
    We qualitatively compare our reconstruction with ground truth and other state-of-the-art techniques. We show the top-left view of $t_0$ and EPI images from three consecutive \gls{lf} frames ($t_0,t_1,t_2$).
    As can be clearly seen from the EPI images, our technique consistently provides accurate reconstructions.
    }
    \label{fig:qual}
\end{figure}

\cref{tab:quant} details the quantitative comparisons of various algorithms
against all $4$ datasets: Hybrid, ViewSynth, TAMULF and Stanford.
Our proposed reconstruction outperforms previous state-of-the-art techniques.
We also notice that even our image-based model `Proposed(image)' outperforms the previous image-based \gls{lf} prediction techniques.
We can also see clear distinction when we compare the images qualitatively in \cref{fig:qual}, especially when the EPI for the \gls{lf} views are taken into account.
We also validate our algorithm on monocular videos acquired from a commercial GoPro camera.
While we show some results from GoPro dataset in \cref{fig:expt:occlusion}, please refer to the supplementary material for more qualitative results.

\begin{table}[t]
\centering
\scriptsize
\setlength\tabcolsep{2pt}
\begin{tabular}{@{}lccccccccccc@{}}
\toprule
\multirow{2}{*}{Algorithm}&\phantom{~} & \multicolumn{2}{c}{Hybrid} & \multicolumn{2}{c}{ViewSynth} & \multicolumn{2}{c}{TAMULF} & \multicolumn{2}{c}{Stanford} & \multicolumn{2}{c}{Average} \\
  & & PSNR & SSIM & PSNR & SSIM & PSNR & SSIM & PSNR & SSIM & PSNR & SSIM \\
\hline
Niklaus \textit{et al.}\cite{niklaus20193d} & & 23.87 & 0.873 & 23.19 & 0.903 & 18.12 & 0.811 & 23.19 & 0.892 & 22.10 & 0.870 \\
Srinivasan \textit{et al.}\cite{srinivasan2017learning} & & 28.12 & 0.893 & 28.56 & 0.931 & 22.63 & 0.857 & 29.24 & 0.924 & 27.14 & 0.901 \\
Li \textit{et al.}\cite{li2020synthesizing} & & 31.62 & 0.950 & 29.39 & 0.945 & 25.63 & 0.903 & 30.44 & 0.956 & 29.27 & 0.938 \\
Li+Ranftl\cite{ranftl2021vision} & & 31.69 & 0.950 & 29.90 & 0.953 & 25.83 & 0.906 & 31.21 & 0.962 & 29.66 & 0.943 \\
Proposed(image) & & \bfg{32.48} & \bfg{0.951} & \bfg{30.76} & \bfg{0.955} & \bfg{27.42} & \bfg{0.927} & \bfg{34.53} & \bfg{0.970} & \bfg{31.30} & \bfg{0.951}\\
Proposed & & \bfb{32.66} & \bfb{0.952} & \bfb{30.97} & \bfb{0.956} & \bfb{27.24} & \bfb{0.922} & \bfb{34.98} & \bfb{0.974} & \bfb{31.47} & \bfb{0.951} \\

\bottomrule
\end{tabular}
\caption{We quantitatively compare our proposed technique with state-of-the-art algorithms on various datasets.
Our algorithm consistently provides high-fidelity reconstructions.
\bfb{Blue} and \bfg{green} represent the top-two performing algorithm in each column.}
\label{tab:quant}
\end{table}

\begin{table}[t]
\centering
\scriptsize
\setlength\tabcolsep{3pt}
\begin{tabular}{@{}lccccccccccc@{}}
\toprule
\multirow{2}{*}{Algorithm}&\phantom{~} & \multicolumn{2}{c}{Hybrid} & \multicolumn{2}{c}{ViewSynth} & \multicolumn{2}{c}{TAMULF} & \multicolumn{2}{c}{Stanford} & \multicolumn{2}{c}{Average} \\
  & & PSNR & SSIM      & PSNR & SSIM & PSNR & SSIM & PSNR & SSIM & PSNR & SSIM \\
\hline
Base & & 30.76 & 0.945 & 29.07 & 0.947 & 25.74 & 0.918 & 31.70 & 0.963 & 29.32 & 0.943 \\
Base+occ & & 31.78 & 0.949 & 29.71 & 0.948 & 26.51 & \bfg{0.919} & 32.99 & 0.965 & 30.25 & 0.945 \\
Base+occ+adpt & & \bfg{32.26} & \bfg{0.950} & \bfg{30.69} & \bfg{0.954} & \bfg{26.96} & \bfg{0.919} & \bfg{34.45} & \bfg{0.973} & \bfg{31.09} & \bfg{0.949} \\
Proposed & & \bfb{32.66} & \bfb{0.952} & \bfb{30.97} & \bfb{0.956} & \bfb{27.24} & \bfb{0.922} & \bfb{34.98} & \bfb{0.974} & \bfb{31.47} & \bfb{0.951} \\

\bottomrule
\end{tabular}
\caption{
We consider a baseline model `Base' that is trained only with the self-supervised constraints as proposed in \stereo \cite{shedligeri2021selfvi}.
We then successively enhance the `Base' model with disocclusion handling, adaptive TD layer and the refinement block and compare the performance boost in each case.
}
\label{tab:ablation}
\end{table}

\subsubsection{Temporal consistency}

We evaluate and quantitatively compare the temporal consistency of the videos predicted from our proposed algorithm.
For this, we first predict optical flow via \cite{teed2020raft} between all \glspl{sai} of successive ground-truth \gls{lf} frames.
We then compute the mean squared error between the current estimated \gls{lf} and the previous \gls{lf} warped to the current frame.
We provide quantitative comparison in the supplementary material.

\subsection{Ablation Study}
\label{sec:ablation}
Our proposed technique contains three key building blocks that enable us to work with monocular videos.
Here, we evaluate the contribution of each of the three building blocks to the reconstruction quality.
As shown in \cref{tab:ablation} we evaluate the effect of each block by successively adding the proposed blocks to the baseline model and quantitatively comparing the reconstructed \gls{lf} videos.
The baseline model can also be thought of as an extension of \stereo \cite{shedligeri2021selfvi} to the case of monocular videos. 
Here, we utilize \emph{only} the geometric, photometric and temporal consistency constraints proposed in \stereo.
The \gls{lf} synthesis network architecture \lfnet remains identical in all the models.

\begin{figure}[t]
    \centering
    \includegraphics[width=\columnwidth]{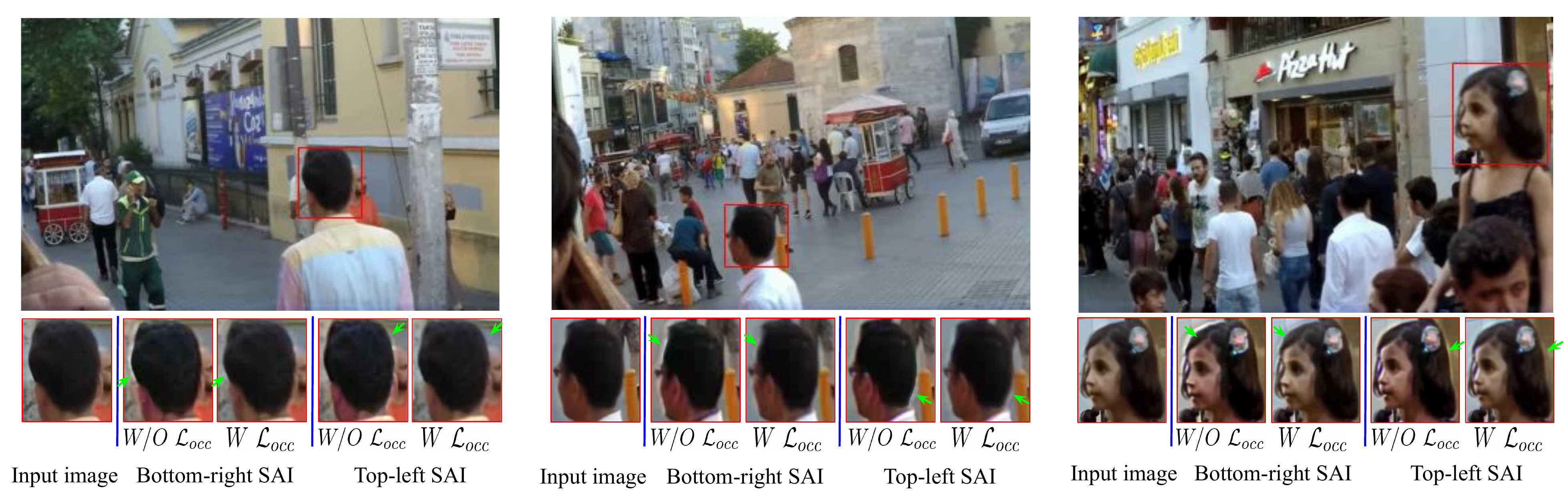}
    \caption{The model trained without disocclusion handling leads to a halo-like artifact around depth-edges in the \glspl{sai} of the frames.
    With the proposed disocclusion handling technique, the network learns to accurately fill-in the disoccluded pixels.
    }
    \label{fig:expt:occlusion}
\end{figure}

\noindent\textbf{Disocclusion handling (Base vs Base+occ):} Enforcing the disocclusion handling constraint helps the synthesis network to learn to fill in the disoccluded pixels in the estimated \gls{lf} frames as shown in \cref{fig:expt:occlusion}.
Quantitatively, we also observe a boost of $0.9$dB PSNR in comparison to the baseline model.

\begin{figure}[t]
    \centering
    \includegraphics[width=\columnwidth]{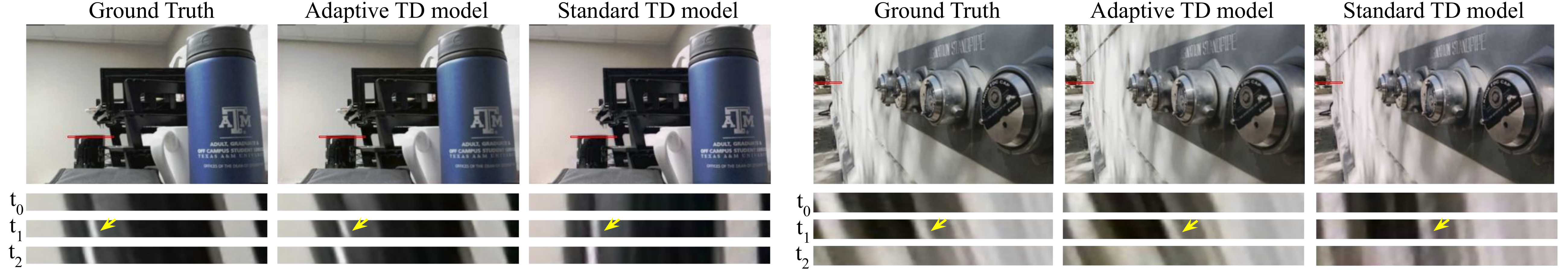}
    \caption{
    As seen in the EPI images, the standard \gls{td} model is unable to represent the depth for the scene accurately compared to the proposed adaptive \gls{td} model.
    By separately determining the depth planes for each scene, adaptive \gls{td} model gives a more accurate reconstruction.
    }
    \label{fig:adaptive_td}
\end{figure}

\noindent\textbf{Adaptive \gls{td} layer (Base+occ vs Base+occ+adpt):}
Our modified adaptive \gls{td} layer can accurately represent the depth planes in the \gls{lf} as can be seen from the EPI images in \cref{fig:adaptive_td}.
Quantitatively, we get a significant performance boost of about $0.7$dB PSNR.

\begin{figure}[t]
    \centering
    \includegraphics[width=\columnwidth]{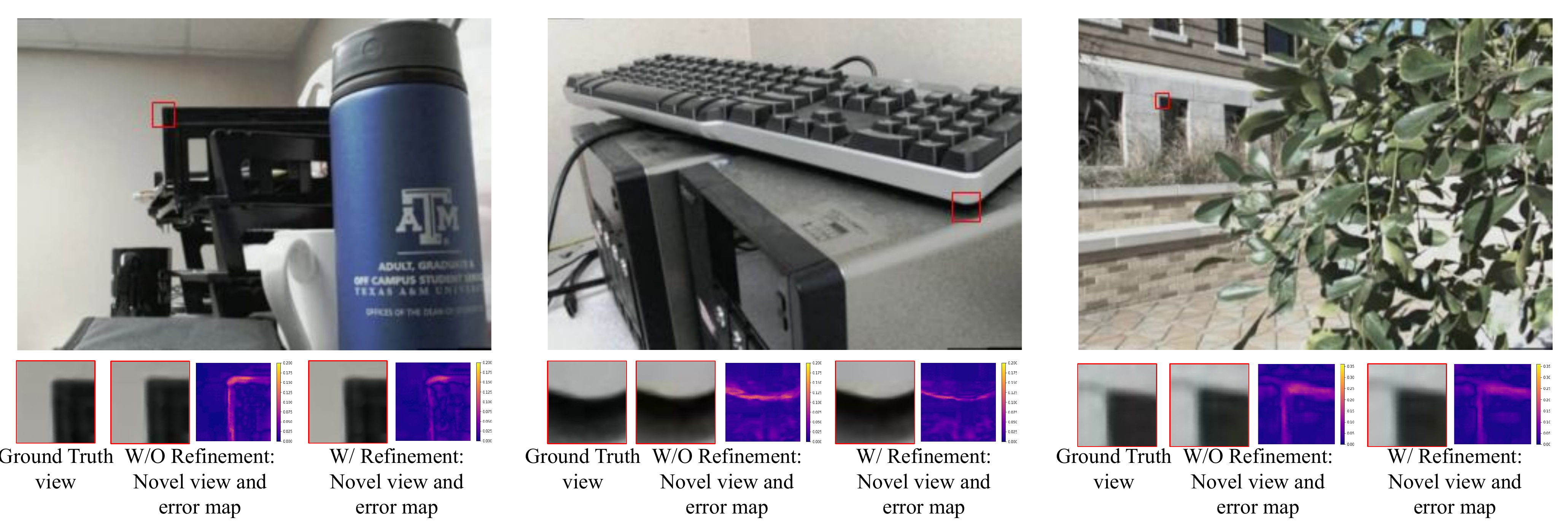}
    \caption{
    The error map between the reconstructed and ground-truth shows that supervised refinement improves reconstruction at depth-edges.
    The refinement module corrects the baseline discrepancy 
    The refinement block utilizes the spatial information in other \gls{sai}s through angular attention and optimizes the positioning of depth-edges correcting the baseline discrepancy between synthesized and ground truth \gls{lf}.
    }
    \label{fig:refinement_qual}
\end{figure}

\noindent\textbf{Supervised refinement block (Base+occ+adpt vs Proposed):}
Finally, we evaluate the effect of the novel refinement block that is trained with supervised loss on ground-truth \gls{lf} frames.
We observe an expected improvement in the reconstruction quality, showing a boost in PSNR of nearly $0.4$dB.
We also make qualitative comparison in \cref{fig:refinement_qual}, where we see that the refinement block provides more accurate \gls{sai}s around depth edges that are difficult to reconstruct with just self-supervised learning.

\subsection{Variable baseline LF prediction}
\label{sec:expt:var_baseline}
Commercial \gls{lf} cameras such as Lytro Illum capture \gls{lf} images with fixed baseline. 
Hence, supervised techniques using this data are also limited and can produce \gls{lf} images with a fixed baseline.
However, our proposed network reconstructs \gls{lf} frames based on the input disparity map.
By scaling the disparity map by a constant factor, we can scale the disparity values input to the network, leading to \gls{lf} prediction with variable baselines.
In \cref{fig:var_baseline} we demonstrate this with $4$ different scale factor for disparity maps, $1\times$, $1.5\times$, $2\times$, $2.5\times$.
Note that our algorithm allows us to generate \gls{sai}s with higher baseline than that of the ground truth frames from Lytro.

\section{Discussion}
\label{sec:discussion}
\begin{figure}[t]
    \centering
    \includegraphics[width=\columnwidth]{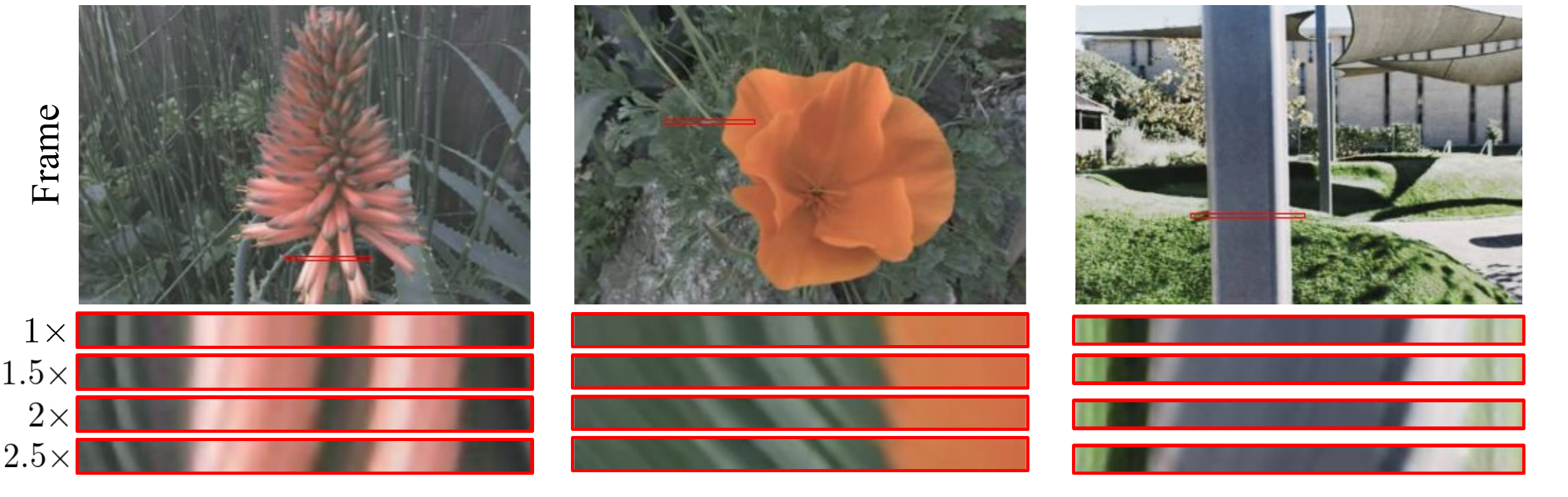}
    \caption{ With our proposed self-supervised technique, \gls{lf} frames with variable baselines can be predicted by just scaling the input disparity map.
    We demonstrate this on $4$ different scales $\{1,1.5,2,2.5\}\times$. Notice the increasing slope in the EPI images from $1\times$ to $2.5\times$. }
    \label{fig:var_baseline}
\end{figure}

Our proposed algorithm is largely a self-supervised technique except for the refinement block that is supervised using ground-truth \gls{lf} \emph{image} data.
The refinement block uses a transformer module for angular attention.
To the best of our knowledge this is also the first attempt to employ vision transformers to \gls{lf} data.
Note that our proposed algorithm outperforms previous state-of-the-art techniques even without the supervised refinement module.
Another point to note is that, during inference, we do not have any information about the true baseline of the \gls{lf}.
We only have access to a relative depth map obtained from a single input image.
Hence, it becomes difficult to accurately compare with the ground-truth.
To solve this, we choose a scale and shift factor ($\{a,b\}$) such that the mean error between the computed disparity maps (from relative depth maps) and the true disparity maps (for a given dataset such as TAMULF) is minimum.
Outside of comparison with ground truth, the true disparity map is not necessary and we can generate \gls{lf} of multiple baselines as needed (\cref{fig:var_baseline}).

\section{Conclusion}
\label{sec:conclusion}
We propose an algorithm for \gls{lf} video reconstruction from just a monocular video input.
Our baseline model utilizes the intermediate low-rank representation for \gls{lf} and the self-supervised geometric, photometric and temporal constraints inspired from \cite{shedligeri2021selfvi}.
However, necessary key modifications were proposed in this work that enabled the final model to reconstruct high-fidelity \gls{lf} videos from monocular input.
We propose a disoclussion handling technique that is required to fill-in disoccluded regions in the estimated \gls{lf}.
We also propose a modified low-rank representation that can adapt to each input scene based on the layer displacements predicted by the network.
Finally, we introduce a novel supervised, transformer-based refinement block that can further refine the predicted \gls{lf} quality.
While showing superior performance with respect to the previous state-of-the-art techniques, our model also enables prediction of \gls{lf} frames with varying baselines.
Overall, our proposed algorithm facilitates a monocular camera for applications like refocusing and novel view synthesis.

\subsubsection*{Acknowledgements}
This work was supported in part by Qualcomm Innovation Fellowship (QIF) India 2021.

{
\bibliographystyle{splncs04}
\bibliography{egbib,bib_report}
}

\appendix

\section{More Details Regarding Experimental Comparisons Performed}
Our proposed self-supervised algorithm is designed to take a monocular video as input and output a \gls{lf} video sequence.
We evaluate our algorithm against other state-of-the-art supervised monocular \gls{lf} estimation algorithms \cite{li2020synthesizing,srinivasan2017learning,niklaus20193d}.
For Li \textit{et al.} \cite{li2020synthesizing}, we use their publicly available implementation to make our comparisons.
To obtain the complete \gls{lf} video from \cite{li2020synthesizing}, we have to reconstruct each frame of the video individually.
Li \textit{et al.} + Ranftl \textit{et al.} represents a modified version of \cite{li2020synthesizing}, where we input a depth estimate from DPT \cite{ranftl2021vision} instead of the original DeepLens \cite{DBLP:journals/corr/abs-1810-08100} model.
Since \cite{li2020synthesizing} is not trained on depth inputs from DPT \cite{ranftl2021vision}, we finetune \cite{li2020synthesizing} on the TAMULF dataset with depth maps obtained from DPT \cite{ranftl2021vision}.
For finetuning, we use AdamW optimizer for $5$ epochs, with an initial learning rate of $2\times 10^{-5}$ and a weight decay of $0.001$.
While \cite{srinivasan2017learning} is originally  trained on a dataset of flower images (proposed in the same work), we finetune it on a larger and diverse TAMULF dataset from \cite{li2020synthesizing}.
This is done because the test data has much more diverse inputs than the images in the original flowers dataset.
The network is finetuned in Tensorflow using Adam optimizer for $80k$ iterations with a learning rate of $10^{-4}$.

In \cref{tab:quant} of main, we perform quantitative comparisons of various algorithms against 4 datasets: Hybrid, ViewSynth, TAMULF and Stanford.
From the \emph{Hybrid} dataset we consider the central $7\times 7$ views as the ground-truth light field videos, and the center-view of each \gls{lf} forms the input monocular video.
The rest three datasets are \gls{lf} \emph{image} datasets, and we simulate \gls{lf} videos with $8$ frames from each \gls{lf} following the procedure described in \cite{shedligeri2021selfvi,lumentut2019deep}.

In case of Hybrid, we use the test sequences from the dataset.
For ViewSynth datasets, we choose the synthesized video sequences from the default test set.
For TAMULF dataset, we randomly chose $84$ samples from $1084$ light field frames, and these frames are used to  synthesize videos which are used for testing the algorithms.
The Stanford dataset contains $4211$ light fields organised into 30 categories. 
Each scene is captured from $3, 4$ or $5$ different camera poses, over a total of $850$ scenes. 
We randomly select $113$ light fields frames from \textit{Tree} category to synthesize videos which was used as the test set.

\section{Details of our Proposed Network architecture}
\begin{figure}[t]
    \centering
    \includegraphics[width=\textwidth]{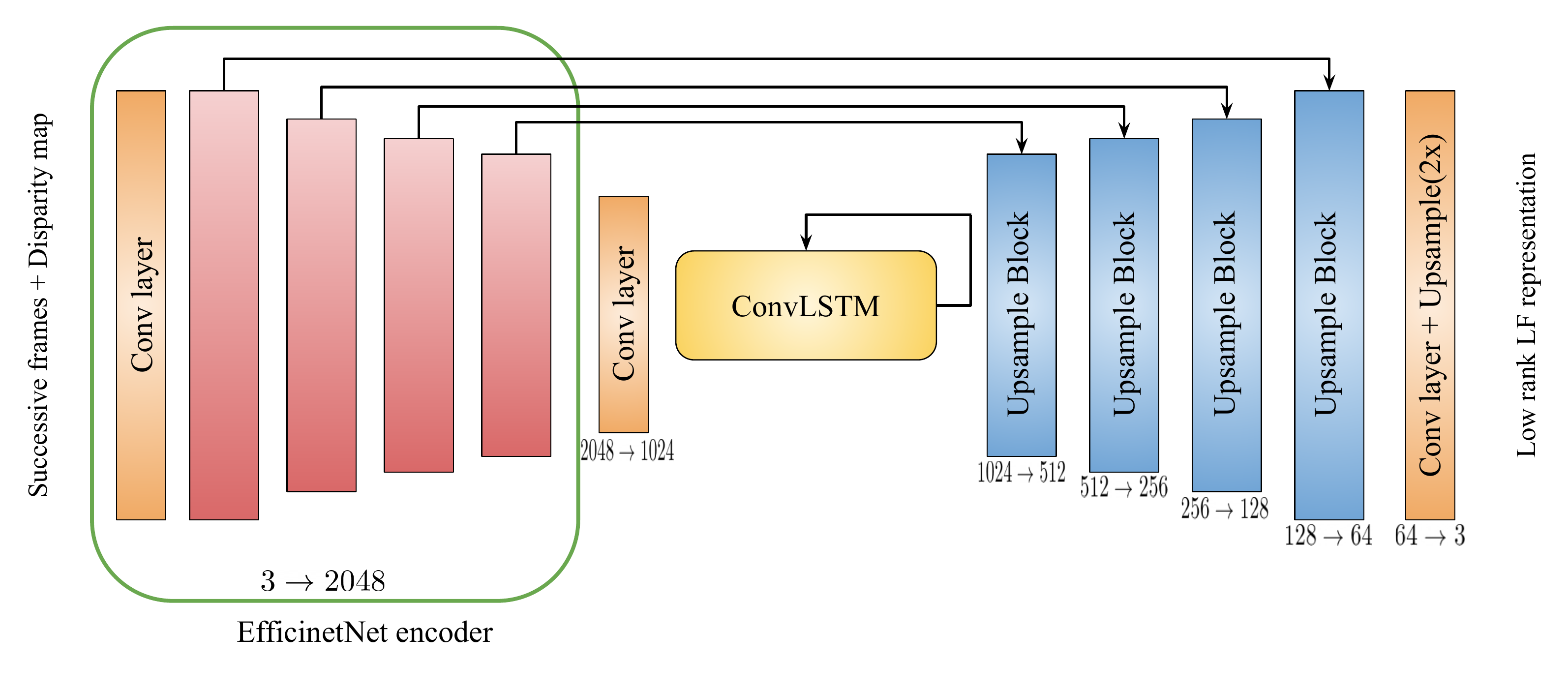}
    \caption{
    We show the detailed network architecture of the \gls{lf} synthesis network {\lfnet} that predicts the low-rank representation \lowrank.
    For estimation of the low-rank \gls{lf} representation, we use a encoder-decoder based network with ConvLSTM module used for learning temporal information.
    The encoder block follows the Efficient-Net\cite{tan2019efficientnet} architecture and the decoder consists of bilinear interpolation operation followed by convolution and batch-normalization operation.
    }
    \label{fig:supp:lfnet}
\end{figure}

\paragraph{Light field synthesis network, $\lfnet$} 
The synthesis network \lfnet\ is a \gls{lstm} based recurrent neural network consisting of a Efficient-Net encoder \cite{tan2019efficientnet} and a convolutional decoder with skip connections as shown in \cref{fig:supp:lfnet}.
The LSTM layer follows the Efficient-Net encoder and the cell output from the LSTM layer is fed as input to the decoder network.
The decoder network consists of $4$ upsampling blocks.
In the upsampling block, the feature maps are first doubled in size spatially using bilinear interpolation. 
The upsampled feature map is then fed to a series of two convolutional layers of filter size $3\times 3$, which is then followed by batch normalization.
The output of the final upsampling block is then input to a final convolutional layer which outputs 36 RGB (108) channels.
These $108$ channels correspond to the $N = 3$ layers and $R = 12$ rank of the low-rank \gls{lf} representation \lowrank.

The displacements $D=\{D_1, D_2, D_3\}$ for the Adaptive \td layer are predicted from \gls{mVIT}\cite{bhat2021adabins} network that takes as input  \{\prevframe,\currframe,\nextframe,\currdisp\}.
The \gls{mVIT} network used in our work is exactly the same as the one proposed in \cite{bhat2021adabins}.
Except that in our implementation we stack successive frames and disparity map as input ($10$ channels) and provide it as input to \gls{mVIT}.
The output of this network is a sequence of values $D=\{D_1, D_2, D_3\}$ which is used to predict the \gls{lf} frame.

\begin{figure}[ht!]
    \centering
    \includegraphics[width=\textwidth]{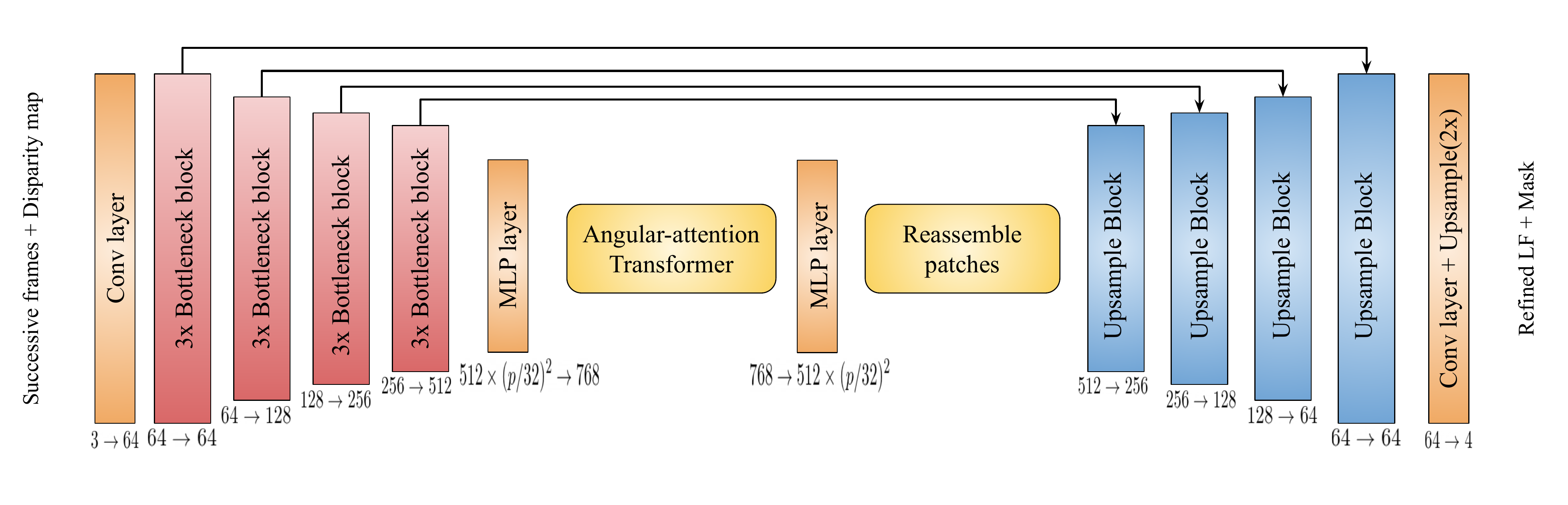}
    \caption{
    We show the detailed architecture of the supervised Refinement module in our proposed algorithm  (see \cref{fig:refinement} of main).
    It takes as input the spatial patches cropped from each \gls{lf} \gls{sai} and produces a feature map for independently for each patch.
    The `Bottleneck block' used here is identical to the ResNet-based bottleneck block proposed in \cite{DBLP:journals/corr/HeZRS15}.
    The output features from encoder are then flattened and passed through a MLP layer to get the feature embeddings.
    The transformer layer performs multi-headed self-attention (MHSA) on the embeddings and outputs tokens which are then reassembled spatially to form larger feature maps (see \cref{sec:refinement} of main).
    These feature maps are input to the decoder block identical to the decoder block in \lfnet.
    The refined \gls{lf} and mask are obtained as outputs from the decoder block.
    }
    \label{fig:supp:refnet}
\end{figure}

\paragraph{Supervised residual refinement module} 
Vision Transformers(ViT) \cite{dosovitskiy2020image} form the backbone of our proposed refinement module.
We divide the predicted \gls{lf} frame \estcurrlf\ into non-overlapping patches, each of size $p\times p$ ($=32\times 32$).
A shallow ResNet-based encoder (see \cref{fig:supp:refnet}) extracts features independently from each of the $U^2(=49)$ patches.
The encoder contains $12$ bottleneck blocks \cite{DBLP:journals/corr/HeZRS15} with max-pooling operation carried out at the first of every three blocks as shown in \cref{fig:supp:refnet}.
The obtained feature embeddings are input to two transformer layers which applied  multi-headed self-attention (MHSA) to these tokens.
Here, $P(=6)$ is the number of non-overlapping patches cropped from each angular view.
These $P$ tokens are stacked horizontally and vertically following the order of cropped patches, so as to form a larger feature map.
A shallow decoder network then takes these stacked tokens as input and predicts a $4$ channel output.
The shallow network consists of 4 upsample blocks where each block has the exact same configuration followed in \gls{lf} synthesis network, \lfnet.

\section{Ablation on pre-trained optical flow and depth estimation networks}
Our proposed self-supervised algorithm aims to synthesize \gls{lf} video sequence from monocular video by enforcing geometric consistency via relative depth map, temporal consistency and dis-occlusion handling via optical flow.
For this purpose, we use a pre-trained DPT\cite{ranftl2021vision} network to estimate the relative depth map from monocular frame and pre-trained RAFT\cite{teed2020raft} network to estimate the optical flow between successive input video frames.
Since the performance of our proposed methods depends highly on the pre-trained depth and optical flow network, we compare the effect on performance by replacing DPT and RAFT models with DeepLens\cite{DBLP:journals/corr/abs-1810-08100} and LiteFlowNet\cite{hui18liteflownet} models for relative depth and optical flow estimation.
As shown in \cref{tab:ablation_full}, when the DPT and/or RAFT models are replaced with DeepLens and LiteFlowNet respectively, we observe a decline in the performance of the \gls{lf} synthesis network, \lfnet, as evidenced by the reduction in PSNR and SSIM scores.
From this we conclude that the proposed self-supervised algorithm indeed utilizes the information from the relative depth map and optical flow map and the quality of these estimates affect the performance of the \gls{lf} synthesis network, \lfnet.
Also, our proposed supervised refinement module tries to correct these errors using the available ground-truth data, irrespective of the relative depth and optical flow networks, resulting in similar quality \gls{lf} estimates after refinement.

\begin{table}[t]
\centering
\setlength\tabcolsep{5pt}
\begin{tabular}{@{}lccccl@{}}
\toprule
\multirow{2}{*}{Model} & \multirow{2}{*}{Depth} & \multirow{2}{*}{Flow} & \multirow{2}{*}{Refine} & \multicolumn{2}{c}{Average} \\
   & & & & PSNR & SSIM \\
  \hline
  V1 & DPT & RAFT & \redmark & 31.09 & 0.949 \\
  V2 & DPT & RAFT & \greencheck & 31.47 & 0.951 \\
  V3 & Deeplens & RAFT & \redmark & 30.53 & 0.945 \\
  V4 & Deeplens & RAFT & \greencheck & 30.76 & 0.946 \\
  V5 & DPT & LiteFlow & \redmark & 29.60 & 0.949 \\
  V6 & DPT & LiteFlow & \greencheck & 31.27 & 0.950 \\
  V7 & Deeplens & LiteFlow & \redmark & 29.01 & 0.941 \\
  V8 & Deeplens & LiteFlow & \greencheck & 29.59 & 0.941 \\
\bottomrule
\end{tabular}
\caption{
\textbf{Effect of depth and optical flow accuracy on our proposed method:} Accurate depth and optical flow estimates are necessary to enforce the geometric and temporal consistency on the predicted \gls{lf} video. 
Estimated optical flow is also used to handle disoccluded pixels in the \gls{lf}.
Hence, we compare the effect of replacing these state-of-the-art models (DPT and RAFT) with slightly less accurate models (DeepLens\cite{DBLP:journals/corr/abs-1810-08100} and LiteFlowNet\cite{hui18liteflownet}).
We observe that using less accurate depth and optical flow estimates causes our reconstruction results to degrade.
And this degradation is more pronounced when we don't use the proposed supervised refinement module.
This also points to the conclusion that our proposed model is indeed utilizing the information from the depth and optical flow estimates.
Also, our proposed supervised refinement module tries to correct these errors using the available ground-truth data.
}
\label{tab:ablation_full}
\end{table}

\section{Temporal consistency of synthesized LF video sequence}
Our proposed algorithm aims to reconstruct \gls{lf} \emph{video} sequences where temporal consistency is a crucial factor.
We evaluate and quantitatively compare the temporal consistency of the videos predicted from our proposed algorithm.
For evaluating temporal consistency between successive predicted \gls{lf} frames, we first predict  optical flow via \cite{teed2020raft} between all \glspl{sai} of successive ground-truth \gls{lf} frames, i.e., \truelf, \prevtruelf.
The current estimated \gls{lf} is then warped to the previous \gls{lf} frame using the estimated ground-truth optical flow.
We then compute the mean squared error between the previous estimated \gls{lf} and the current \gls{lf} warped to the previous \gls{lf}.
The warping error is used as a measure of temporal stability between successive predicted \gls{lf} frames.
We calculate the temporal stability function (lower is better) as 
\begin{equation}
    \tempstab(\estcurrlf; \truelf, \prevtruelf) = \sum_{\mathbf{u}} \| \warp\left(\estcurrlf\left(\mathbf{u}\right) ; \flownet(\truelf\left(\mathbf{u}\right), \prevtruelf\left(\mathbf{u}\right)) \right) - \prevtruelf \|_2
    \label{eq:temp_stab}
\end{equation}

To estimate the temporal stability of the network for a video, the {\tempstab} function is then averaged over the entire video.
In table \cref{tab:temp_stab}, we compare two of our models, `Base+occ+adpt' (without refinement block) and `Proposed' (with refinement block), with previous learning-based techniques.
We observe that our proposed algorithm performs significantly better than previous state-of-the-art techniques for \gls{lf} estimation.
Although the `Proposed' model is trained on images without the temporal consistency constraint, its performance does not degrade compared to `Base+occ+adpt' model which is trained on monocular videos with explicit temporal consistency loss.

\begin{table}[t]
\centering
\setlength\tabcolsep{3pt}
\begin{tabular}{@{}lcccccccccccc@{}}
\toprule
Algorithm & \phantom{a} & Hybrid & ViewSynth & TAMULF & Stanford & Average \\
\hline

Niklaus & & 0.357 & 0.070 & 0.219 & 0.065 & 0.177 \\
Srinivasan & & 0.195 & 0.020 & 0.070 & 0.014 & 0.075\\
Li & & 0.108 & 0.019 & 0.034 & 0.009 & 0.043 \\
Li+Ranftl & & 0.108 & \bfb{0.016} & 0.033 & 0.008 & 0.042 \\
Base+occ+adpt & & \bfg{0.103} & 0.017 & \bfg{0.028} & \bfg{0.006} & \bfg{0.038}\\
Proposed & & \bfb{0.102} & \bfg{0.016} & \bfb{0.027} & \bfb{0.006} & \bfb{0.038}\\

\bottomrule
\end{tabular}
\caption{\textbf{Evaluating temporal consistency:} We quantitatively compare the temporal consistency of the predicted \gls{lf} videos through checking the consistency with the optical flow.
Our proposed algorithm performs significantly better than previous state-of-the-art techniques for \gls{lf} estimation.
Although the refinement block of the `Proposed' model is trained on images without the temporal consistency constraint, its performance does not degrade compared to `Base+occ+adpt' model which is trained on monocular videos with explicit temporal consistency loss.
\bfb{Blue} and \bfg{green} represent the top-two performing algorithm in each column.
}
\label{tab:temp_stab}
\end{table}

\section{Identified depth planes for adaptive tensor-display model}
\begin{figure}[t]
    \centering
    \includegraphics[width=0.99\textwidth]{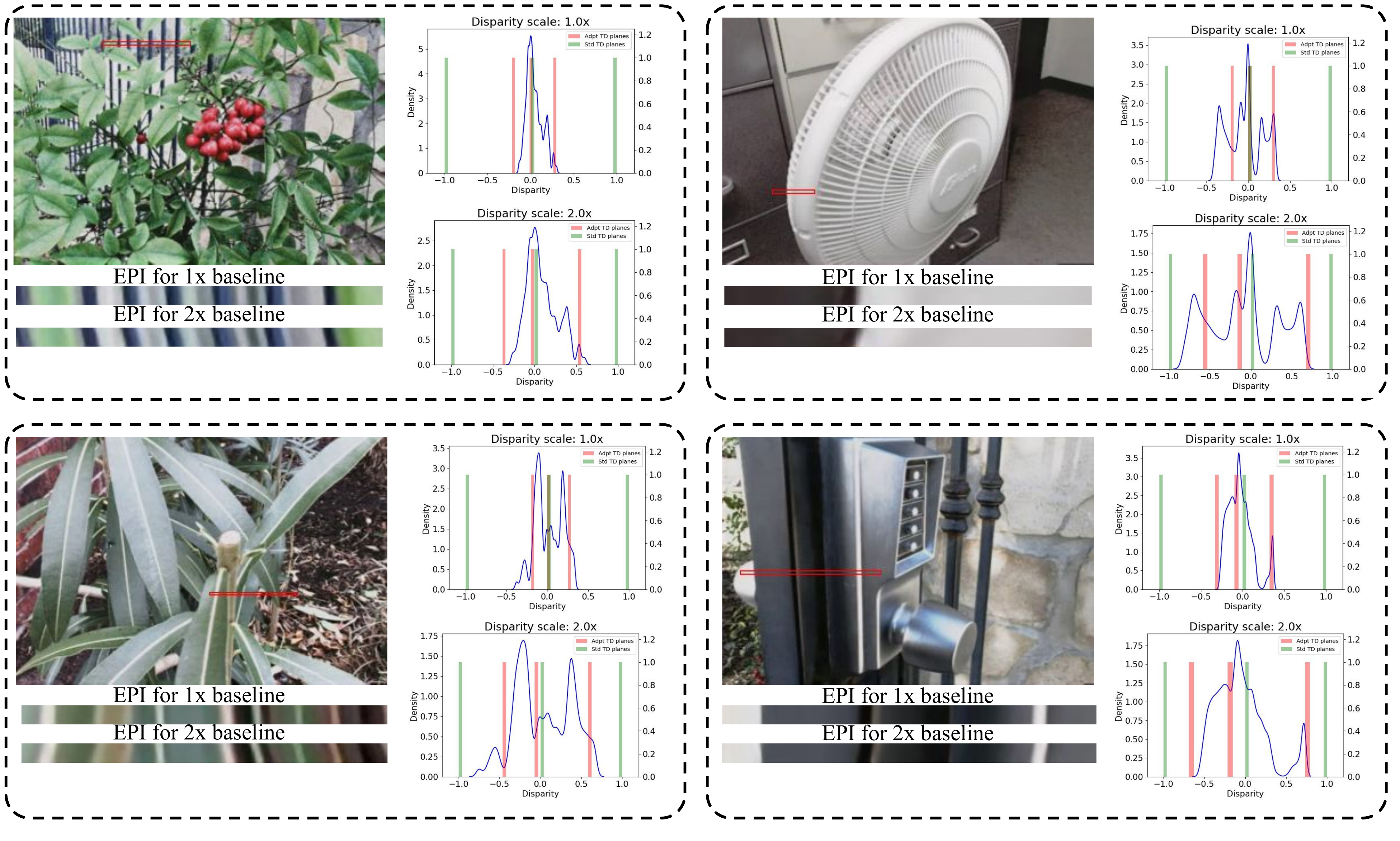}
    \caption{\textbf{Adaptive vs. vanilla tensor-display representation for \glspl{lf}:}
    We show four different samples with EPIs for \gls{lf} predicted by $1\times$ and $2\times$ scaled disparity inputs.
    Besides each figure we also show the distribution of disparity values in the $1\times$ and $2\times$ scaled disparity maps (shown as \bluecolor{blue} curve).
    In these graphs we also represent the disparity planes in the standard and adaptive TD models with green and orange bars respectively.
    We observe that the green bars remain constant (at $-1,0,+1$) even when the disparity gets scaled.
    While the predicted orange bars adapt to the input scaled disparity maps thereby providing a more accurate representation of the \gls{lf}.
    We also notice that the $3$ orange bars are not necessarily uniformly distant from each other.
    Refer to supplementary video file for video results.
    }
    \label{fig:supp:adpt-td}
\end{figure}

We propose an adaptive tensor-display based low rank representation for estimating \gls{lf} from the given input frames.
Here, we showcase how the distance values $D=\{D_1, D_2, D_3\}$ adapt to the varying input depth maps, thereby providing superior reconstruction results than that of the standard tensor-display based low-rank representation.
In \cref{fig:supp:adpt-td}, we show various \gls{lf} images predicted by different scaled versions of the disparity map.
Specifically, we scale the input disparity map by the factors $1$ and $2$ and predict the corresponding \gls{lf}.
We observe that the predicted depth planes adapt to the input disparity maps providing superior reconstruction results.

\section{Application to video refocusing}
\begin{figure}[t!]
    \centering
    \includegraphics[width=\columnwidth]{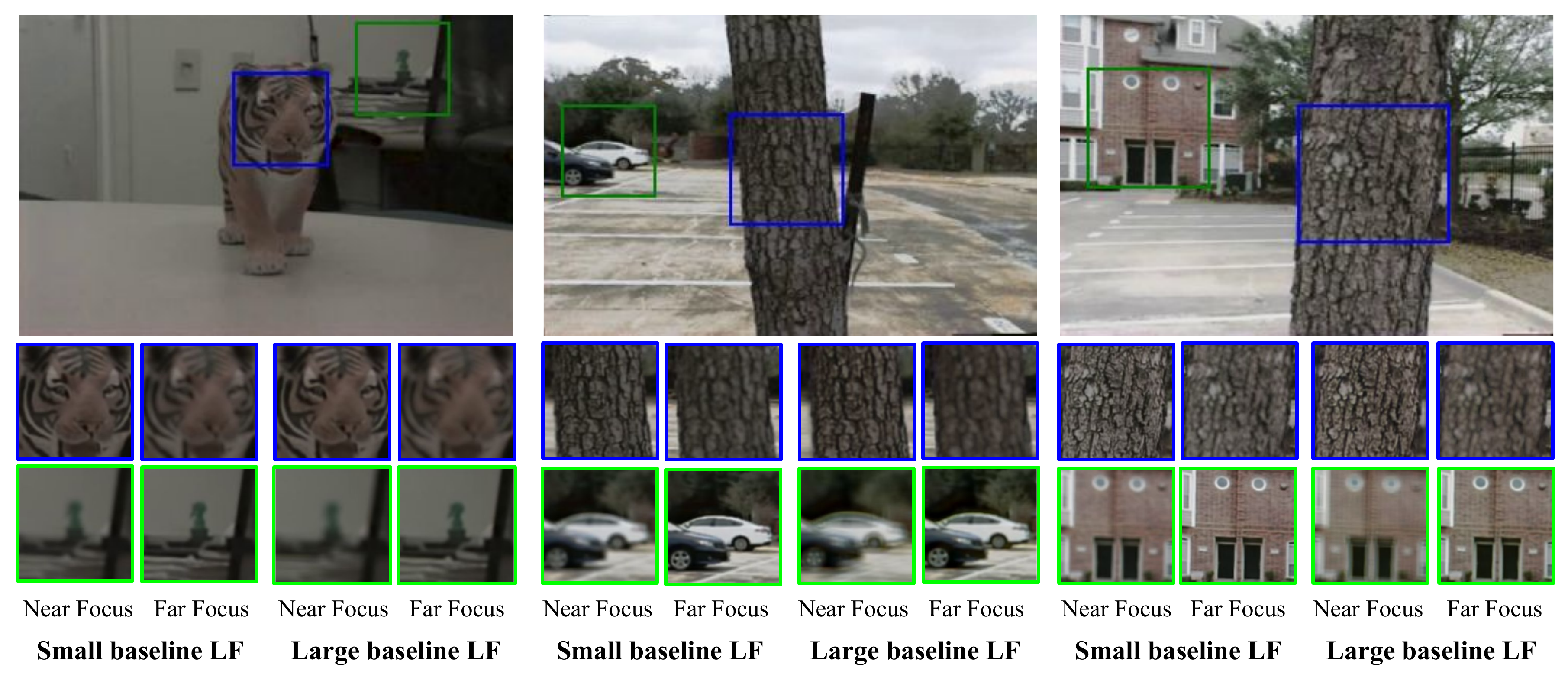}
    \caption{As our model can output \glspl{lf} with varying baselines, we can demonstrate refocusing effects with varying blur sizes.
    As can be seen, the blur size in a large baseline \gls{lf} is bigger than the one in a small baseline \gls{lf}.
    }
    \label{fig:supp:refocusing}
\end{figure}

\gls{lf} have been popular because they provide a very intuitive way of doing post-capture focus control.
The amount of defocus that can be achieved depends on the baseline of the \gls{lf}.
As demonstrated in \cref{sec:expt:var_baseline}, our technique is not limited to a single baseline.
Unlike previous \gls{lf} prediction techniques, a single model can output \gls{lf} frames with multiple baselines. And this can be controlled by simply increasing/decreasing the scale factor used to convert a relative depth map to disparity map.
As shown in \cref{fig:supp:refocusing}, this can be used to control the level of blur in the defocused region.
In a typical \gls{lf} camera, post-capture aperture control can be used to only reduce the blur size from a maximum position.
Here, by predicting a \gls{lf} with a larger baseline, we can also increase the defocus blur.

\end{document}

%% file: macros.tex
\usepackage[acronym]{glossaries}
\usepackage{stmaryrd}
\makeglossaries
\input{abbreviations}
\usepackage{pifont}
\usepackage{balance}
\usepackage[dvipsnames,table,xcdraw]{xcolor}
\newcommand{\bluecolor}[1]{\textcolor{blue}{#1}}

\newcommand{\currframe}{\ensuremath{I_t}}
\newcommand{\prevframe}{\ensuremath{I_{t-1}}}
\newcommand{\nextframe}{\ensuremath{I_{t+1}}}
\newcommand{\currdisp}{\ensuremath{d_t}}
\newcommand{\currdepth}{\ensuremath{z_t}}

\newcommand{\loss}{\ensuremath{\mathcal L}}
\newcommand{\warp}{\ensuremath{\mathcal W}}
\newcommand{\ford}{$4$D }
\newcommand{\td}{TD }

\newcommand{\lfnet}{\ensuremath{\mathcal V~}}
\newcommand{\flownet}{\ensuremath{\mathcal O}}

\newcommand{\estcurrlf}{\ensuremath{\mathbf{\hat L}_t}}
\newcommand{\estcurrlfref}{\ensuremath{\mathbf{\hat L}_t^{ref}}}
\newcommand{\outputref}{\ensuremath{\mathbf{\tilde L}_t^{ref}}}
\newcommand{\truelf}{\ensuremath{\mathbf{L}_t}}
\newcommand{\prevtruelf}{\ensuremath{\mathbf{L}_{t-1}}}

\newcommand{\bfu}{\ensuremath{\mathbf{u}}}
\newcommand{\lowrank}{\ensuremath{\mathbf{\mathcal{F}}}}

\newcommand{\bfg}[1]{\textcolor{ForestGreen}{\textbf{#1}}}
\newcommand{\bfb}[1]{\textcolor{blue}{\textbf{#1}}}

\newcommand{\currmask}{\ensuremath{\mathcal{M}_t}}
\newcommand{\frwdwarpprevframelf}{\ensuremath{\tilde{\bf  L }_{t\shortleftarrow t-1}(\bfu)}}
\newcommand{\frwdwarpnextframelf}{\ensuremath{\tilde{\bf L }_{t\shortleftarrow t+1}(\bfu)}}

\newcommand{\stereo}{SeLFVi}

\newcommand{\tempstab}{\ensuremath{\mathcal E_{temp}}}

\newcommand*\colourcheck[1]{%
  \expandafter\newcommand\csname #1check\endcsname{\textcolor{ForestGreen}{\ding{52}}}%
}

\newcommand*\colourmark[1]{%
  \expandafter\newcommand\csname #1mark\endcsname{\textcolor{#1}{\ding{55}}}%
}
\colourmark{red}
\colourcheck{blue}
\colourcheck{green}
\colourcheck{red}


%% file: abbreviations.tex
\newacronym{lf}{LF}{light-field}
\newacronym{fps}{fps}{frames per second}
\newacronym{mp}{MP}{megapixels}
\newacronym{mpi}{MPI}{Multi-Plane Image}
\newacronym{mVIT}{m-VIT}{mini-vision transformer}
\newacronym{vit}{ViT}{vision transformer}
\newacronym{sai}{SAI}{sub-aperture image}
\newacronym{lstm}{LSTM}{long short term memory}
\newacronym{mhsa}{MHSA}{multi-headed self-attention}
\newacronym{td}{TD}{tensor-display}